\newcommand{\beq}{\begin{equation}}
\newcommand{\eeq}{\end{equation}}
\newcommand{\bqa}{\begin{eqnarray}}
\newcommand{\eqa}{\end{eqnarray}}
\documentclass[12pt]{article}
\usepackage{epsfig}
\def\square{\vcenter{\vbox{\hrule height.4pt
          \hbox{\vrule width.4pt height8pt
          \kern8pt\vrule width.4pt}\hrule height.4pt}}}

\begin{document}

\begin{flushleft}\hspace{12cm}
OHSTPY-HEP-T-98 \\ \hspace{12cm}
hep-ph/9804280 \\ \hspace{12cm}
\today  \\
\end{flushleft}

\vskip 10mm

\centerline{\Large\bf Dimensional Reduction of the Two Higgs}
\centerline{\Large\bf Doublet Model at High Temperature}
\vskip 10mm
\centerline{Jens O. Andersen}
\centerline{\it Department of Physics, The Ohio State University,
Columbus, OH 43210}
\vskip 3mm

\begin{abstract}
{\footnotesize 
Dimensional reduction and effective field theory methods are applied
to the Two Higgs Doublet Model at finite temperature.
A sequence of two 
effective three-dimensional field theories 
which are valid on successively longer distance scales is constructed.
The resulting Lagrangian can be used to study 
different aspects of the phase 
transition in this model as well as the sphaleron rate immediately after
the phase transition.}
\end{abstract}
\pagebreak
\section{Introduction}
The electroweak phase transition (EWPT)
has been the subject of intense investigation
in recent years, largely due to its possible role in generating the 
baryon asymmetry of the present Universe~[1-2]
(see also~\cite{trodden} for a detailed review).
If the electroweak phase transition is of first order, it proceeds
through bubble nucleation, and the baryon asymmetry is produced as
the bubbles expand and the Universe is far from equilibrium [1-2].
Moreover, if the baryon asymmetry produced during the
phase transition has survived until today, 
baryon number violating processes (sphaleron processes) 
must have been suppressed immediately after the phase 
transition~\cite{cohen}.
This requires that the electroweak phase transition is strongly
first order~\cite{cohen}.
It is now well established that this requirement is not met in the 
Standard Model (SM); for realistic values of the Higgs mass, 
the phase transition is either too weakly first order  
to suppress the sphaleron processes, or it is
second order, or there is no phase transition at all~\cite{nope}. 
Furthermore, it is not clear that the amount of CP violation in the
SM  is sufficient.

The fact that baryogensis is ruled out in the Standard Model
suggests the investigation of extensions of the 
Standard Model such as the Minimal Supersymmetric Standard Model (MSSM)
and the Two Higgs Doublet Model (2HDM)~\cite{hunter}. 
Both these theories have additional sources of CP violation.
The main objective is to find
regions in parameter space  where the phase transition is 
strongly first order, so that the excess of baryons produced during the phase
transition is not washed out by sphaleron processes
immediately after the phase transition. 

The electroweak phase transition in the Standard Model~[7-9]
as well as in the Two Higgs Doublet Model~[10-14] 
has been studied using resummed perturbation
theory. The strength of the EWPT in the 
Standard Model weakens as the Higgs mass
increases~[7-9]. However, the resummed loop expansion 
breaks down for large (realistic) Higgs masses (see e.g. Ref.~\cite{arnold12}
for a discussion of the validity of the resummed
perturbation expansion), 
and so one 
must employ nonperturbative methods in order to discriminate between
a weakly first order and a second order phase transition.

The electroweak phase transition in the SM
has also been investigated 
by lattice simulations directly in four dimensions~[15-17], renormalization
group techniques~[18-19] and the $\epsilon$-expansion~\cite{epsilon}.
These methods yield results for the 
quantities characterizing the phase transition that
are in qualitative agreement with the
perturbative treatment.

Significant progress in the study of phase transitions has been made 
by applying the methods of dimensional reduction~[21-24] 
and effective field theory~\cite{lepage}. 
The idea is to integrate out the nonzero bosonic modes as well
as the fermionic modes which decouple from the static modes at high
temperature~[21-24]. One is then left with an effective three-dimensional
field theory of the zero modes. Since the nonstatic modes have masses
of order $T$, the process of dimensional reduction is free of infrared
problems, and in weakly coupled theories this can normally be carried out
perturbatively.
In nonabelian gauge theories the effective three-dimensional theory
contains two momentum scales~\cite{gross}. The scale 
$gT$ is provided by the
temporal component of the gauge field, and the scale $g^2T$
is provided by the spatial components of the gauge field. 
Moreover, in theories with a single Higgs multiplet, 
the scalar mass is normally of order $g^2T$
for temperatures close to 
$T_c$. For theories with more than one Higgs multiplet, the masses of the
additional scalar fields are generally of order $gT$.
In either case
it proves useful to construct a second effective field theory by
integrating out the timelike component of the gauge field
(and possibly some scalar fields)
as shown by Farakos {\it et al.}~\cite{rummu} and by Braaten and 
Nieto~\cite{braaten}. This approach has made effective field theory 
a very powerful tool for studying field theories at high temperatures.
Perturbation theory breaks down for the resulting effective theory 
close to the phase transition and it
is also severely infrared divergent in the symmetric phase.
So one must use nonperturbative methods such as lattice simulations
to investigate the phase transition.
Dimensional reduction has been applied to 
a number of theories with spontaneously broken gauge theories:
$SU(2)$+Higgs~[27,29-30], the Standard Model~\cite{laine}, 
the MSSM~[32-36], $SU(5)$+Higgs~\cite{raja}, 
the 2HDM~\cite{marta} and $U(1)$+Higgs~[27,38-39].
The three-dimensional effective theories have been studied
numerically in e.g. Ref,~[40-44] for $SU(2)$+Higgs, in Ref.~\cite{kaja4} 
in the case of $SU(2)\times U(1)$+Higgs, in Refs. [39,46-53] for $U(1)$+Higgs,
and in Ref.~\cite{su2su3} for $SU(3)\times SU(2)$ with two scalar fields
(The latter arises as an effective $3d$ theory for MSSM for some values
of the parameters). 

In the present paper we reconsider the Two Higgs Doublet Model.
The model is interesting in its own right, but the 2HDM (with the temporal
component of the gauge field as an additional
adjoint Higgs field) in $3d$ 
also arises as an effective theory
for the MSSM~\cite{laine}.
In Ref.~\cite{marta} dimensional reduction for this model was
carried out in the one-loop approximation. The second effective theory
obtained by integrating out the timelike component of the gauge field
was constructed in Refs.~\cite{mikko,marta}, also at the one-loop level.
In order to obtain
a satisfactory accuracy for the thermodynamic quantities
describing the phase transition, 
the scalar mass parameters 
in the $3d$ theory are 
needed to two-loop order~\cite{laine}. This calculation is carried
out in the present paper.
  
The plan of the article is as follows. In section II we 
briefly discuss the Lagrangian for the Two Higgs Doublet Model. 
In section III we present the parameters of the first effective theory.
Section IV is devoted to the scenario where one of the Higgs doublets is
heavy, and is integrated out together with the timelike
component of the gauge fields. In Section V we consider the case where both
Higgs doublets are light and are retained in the final effective Lagrangian.  
Finally, in section VI we summarize.
In Appendix A, the notation and conventions are given. We also list
the necessary sum-integrals in the underlying theory as well as the
three-dimensional integrals needed in the effective theory. In Appendix B, 
some 
details of the matching procedure are given by
explicitly calculating a mass parameter in the first effective theory.
\section{Two Higgs Doublet Model}
The Euclidean Lagrangian for the $SU(2)$ gauge-invariant 2HDM 
without fermions reads
\bqa\nonumber
{\cal L}&=&\frac{1}{4}G_{\mu\nu}G_{\mu\nu}
+(D_{\mu}\Phi_1)^{\dagger}(D_{\mu}\Phi_1)
+(D_{\mu}\Phi_2)^{\dagger}(D_{\mu}\Phi_2)
+m_1^2\Phi_1^{\dagger}\Phi_1+m_2^2\Phi_2^{\dagger}\Phi_2
\\&&
\null +m_3^2(\Phi_1^{\dagger}\Phi_2+\Phi_2^{\dagger}\Phi_1)
+V(\Phi_1,\Phi_2)\, .
\eqa
Here, $\Phi_1$ and $\Phi_2$ are the $SU(2)$-doublets
\beq
\Phi_1=\frac{1}{\sqrt{2}}
\left(\begin{array}{c}
\phi_1+i\eta_1\\
\phi_2+i\eta_2\\
\end{array}\right)\, ,\hspace{1cm}
\Phi_2=\frac{1}{\sqrt{2}}
\left(\begin{array}{c}
\phi_3+i\eta_3\\
\phi_4+i\eta_4\\
\end{array}\right)\, .
\eeq
and
\beq
D_{\mu}\Phi_i=\left(
\partial_{\mu}-ig\tau^{a}A_{\mu}^{a}/2\right)\Phi_i\, , 
\hspace{1cm}G_{\mu\nu}^a=\partial_{\mu}A_{\nu}^a-\partial_{\nu}A_{\mu}^a
+g\epsilon^{abc}A_{\mu}^bA_{\nu}^c\, .
\eeq
Here, $g$ is the gauge coupling, $i=1,2$ and $\tau^{1}$, $\tau^{2}$ and $\tau^{3}$ are the three Pauli matrices.\\ \\
The potential $V(\Phi_1,\Phi_2)$ is~\cite{nope,mikko,marta}
\bqa\nonumber
V(\Phi_1,\Phi_2)&=&\lambda_1(\Phi^{\dagger}_1\Phi_1)^2+\lambda_2(\Phi^{\dagger}_2\Phi_2)^2
+\lambda_3(\Phi^{\dagger}_1\Phi_1)(\Phi^{\dagger}_2\Phi_2)
+\lambda_4(\Phi^{\dagger}_1\Phi_2)(\Phi^{\dagger}_2\Phi_1)\\ \nonumber
&&
+\lambda_5\left[(\Phi^{\dagger}_1\Phi_2)(\Phi^{\dagger}_1\Phi_2)
+(\Phi^{\dagger}_2\Phi_1)(\Phi^{\dagger}_2\Phi_1)\right]
+\lambda_6\left[(\Phi^{\dagger}_1\Phi_1)(\Phi^{\dagger}_2\Phi_1)
+(\Phi^{\dagger}_1\Phi_1)(\Phi^{\dagger}_1\Phi_2)\right] \\
\label{potdef}
&&+\lambda_7
\left[(\Phi^{\dagger}_2\Phi_2)(\Phi^{\dagger}_2\Phi_1)
+(\Phi^{\dagger}_2\Phi_2)(\Phi^{\dagger}_1\Phi_2)\right]\, ,
\eqa
where the scalar self-couplings  are denoted by $\lambda_1-\lambda_7$.
All calculations in the present paper are carried in the Landau gauge.
This is merely a convenient choice, since many diagrams vanish in this
gauge. We emphasize that the parameters of the effective Lagrangians
are gauge fixing independent.
\section{Dimensional Reduction}
In this section we carry out the dimensional reduction step for the 
Two Higgs Doublet Model. 

The fields in the effective Lagrangian are identified 
(up to normalizations) with the zero-frequency
modes of the fields in the full theory. 
If the fields in the effective theory are denoted by
$\Phi_1^{\prime}$, $\Phi_2^{\prime}$, $A_i^{\prime}$ and $A_0^{a \prime}$,
we can schematically write at leading order
\beq
\label{rel}
\Phi^{\prime}_i({\bf x})\approx\sqrt{T}
\int_0^{\beta}d\tau\Phi_i({\bf x},\tau)\,,\hspace{0.35cm}
A_i^{a \prime}({\bf x})\approx\sqrt{T}\int_0^{\beta}
d\tau A_i^a({\bf x},\tau)\,,\hspace{0.35cm}
A_0^{a \prime}({\bf x})\approx\sqrt{T}\int_0^{\beta}d\tau 
A_0^a({\bf x},\tau)\,.
\eeq
The effective Lagrangian consists
of all terms which can built out of the fields
$\Phi_1^{\prime}$, $\Phi_2^{\prime}$, $A_i^{\prime}$ and $A_0^{a \prime}$
and which satisfy the symmetries
present at high temperature. Examples of symmetries are three-dimensional
gauge invariance and an $O(3)$ symmetry for the field $A_0^{a \prime}$.
The effective Lagrangian then reads
\bqa\nonumber
{\cal L}_{\mbox {\scriptsize eff}}^{\prime}&=&\frac{1}{4}G_{ij}^{\prime}
G_{ij}^{\prime}+(D_i\Phi_1^{\prime})^{\dagger}(D_i\Phi_1^{\prime})
+(D_i\Phi_2^{\prime})^{\dagger}(D_i\Phi_2^{\prime})
+M_1^2(\mu)\Phi_1^{\prime\dagger}\Phi_1^{\prime}
+M_2^2(\mu)\Phi_2^{\prime\dagger}\Phi^{\prime}_2 \\ \nonumber
&&
\null +M_3^2(\Phi_1^{\dagger}\Phi_2+\Phi_2^{\dagger}\Phi_1) 
+V(\Phi_1^{\prime},\Phi_2^{\prime})+\frac{1}{2}(D_iA_0^{a \prime})^2
+\frac{1}{2}m_E^2(\mu)(A_0^{a \prime})^2\\
&&
\null+{1\over24}\Lambda_E(\mu)(A_0^{a \prime}A_0^{a \prime})^2
\label{effprim}
+h_E^2(\mu)\Phi_1^{\prime\dagger}\Phi_1^{\prime}A_0^{a \prime}A_0^{a \prime}
+h_E^2(\mu)\Phi_2^{\prime\dagger}\Phi_2^{\prime}A_0^{a \prime}A_0^{a \prime}
+\delta{\cal L}_{\mbox {\scriptsize eff}}^{\prime}\, .
\eqa
Here, we have explicitly written the superrenormalizable part of the
Lagrangian, while $\delta {\cal L}_{\mbox {\scriptsize eff}}^{\prime}$ 
represents all higher order operators
consistent with the
symmetries of the Lagrangian.
The gauge coupling is denoted by $g_{\mbox{\scriptsize E}}^2(\mu)$, the quartic coupling constants 
in the potential $V(\Phi_1^{\prime},\Phi_2^{\prime})$ 
are denoted by $\Lambda_i(\mu)$, and
$D_iA_0^a=(\partial_i+g_E\epsilon^{abc}A_i^b)A_0^c$.
The parameters of ${\cal L}^{\prime}_{\mbox{\scriptsize eff}}$
encode the physics at the scale $T$ and are called short-distance coefficients.
The coefficients of the effective Lagrangian are determined by calculating 
static correlators in the full theory and calculating the corresponding
correlators in the effective theory and require that they be equal at 
distances $R\gg1/T$~[27-28]. 

The matching procedure is complicated by the breakdown of the simple
relations~(\ref{rel}). Beyond leading order we must allow for short-distance
coefficients multiplying the fields in the effective theory. 
At the one-loop level (next-to-leading order), the short-distance
coefficients are given by the momentum dependent part of the two-point
functions, and are associated with field renormalizations in the 
underlying theory. These parameters are called {\it field normalization
constants}, and are denoted by $\Sigma^{(1)\prime}(0)$,
$\Pi_{00}^{(1)\prime}$ and $\Pi^{(1)\prime}$ for the Higgs fields,
the timelike component of the gauge field
and the spatial components of the gauge field, respectively.
The relation between the fundamental scalar field $\Phi_i$ and the scalar
field $\Phi^{\prime}_i$ 
in the effective theory can then schematically be written as
\bqa
\label{relmod}
\left[1-\Sigma^{(1)\prime}(0)\right]^{1/2}
\Phi_i^{\prime}({\bf x})&\approx&\sqrt{T}
\int_0^{\beta}d\tau\Phi_i({\bf x},\tau),
\eqa
and similarly for the other fields. The above remarks also apply when
we consider the two effective three-dimensional field theories in the next
section.

The field normalization constants have been calculated and listed
by Kajantie et al. in~\cite{laine}
for the Standard Model with $N$ Higgs doublets. For $N=2$, the results are
\bqa
\Sigma_1^{\prime}(0)=-\frac{9g^2}{64\pi^2}L_b\, ,\hspace{1cm}
\Pi_{00}^{(1)\prime}(0)=-\frac{g^2}{16\pi^2}\left[4L_b-\frac{10}{3}\right]\, , \hspace{1cm}
\Pi^{(1)\prime}(0)=-\frac{g^2}{16\pi^2}\left[4L_b+\frac{2}{3}\right]\, . 
\eqa
The coupling constants of the scalar fields
have been calculated by Losada in \cite{marta}
at the one-loop level. We list the results here for completeness.
\bqa
\Lambda_1(\mu)&=&\lambda_1T-T\left[12\lambda_1^2+\lambda_3^2
+\lambda_3\lambda_4+\frac{1}{2}\lambda_4^2+2\lambda^2_5+6\lambda^2_6
-\frac{9}{2}\lambda_1g^2+\frac{9}{16}g^4\right]
\frac{L_b}{16\pi^2}+\frac{3}{8}\frac{g^4T}{16\pi^2}\, , \\
\Lambda_2(\mu)&=&\lambda_2T
-T\left[12\lambda_2^2+\lambda_3^2+
\lambda_3\lambda_4+\frac{1}{2}\lambda^2_4+2\lambda^2_5+6\lambda^2_7
-\frac{9}{2}\lambda_2g^2+\frac{9}{16}g^4\right]
\frac{L_b}{16\pi^2}+\frac{3}{8}\frac{g^4T}{16\pi^2}\, , \\\nonumber
\Lambda_3(\mu)&=&\lambda_3T
-T\left[6\lambda_1\lambda_3+2\lambda_1\lambda_4+6\lambda_2\lambda_3
+2\lambda_2\lambda_4+2\lambda_3^2+\lambda_4^2
+4\lambda_5^2+2\lambda_6^2+8\lambda_6\lambda_7\right.\\
&&
\left.\null +2\lambda_7^2-\frac{9}{2}\lambda_3g^2+\frac{9}{8}g^4\right]
\frac{L_b}{16\pi^2}+\frac{3}{4}\frac{g^4T}{16\pi^2}\, ,\\ \nonumber
\Lambda_4(\mu)&=&\lambda_4T
-T\left[2\lambda_1\lambda_4+2\lambda_2\lambda_4
+4\lambda_3\lambda_4+2\lambda_4^2
+32\lambda_5^2+5\lambda_6^2
+2\lambda_6\lambda_7+5\lambda_7^2\right.
\\
&&\left.\null -\frac{9}{2}\lambda_4g^2\right]
\frac{L_b}{16\pi^2},\\ \nonumber
\Lambda_5(\mu)&=&\lambda_5T
-T\left[4\lambda_1\lambda_5+4\lambda_2\lambda_5
+8\lambda_3\lambda_5+12\lambda_4\lambda_5+5\lambda^2_6
+2\lambda_6\lambda_7+5\lambda_7^2\right.\\
&&
\left.\null -\frac{9}{2}\lambda_5g^2\right]
\frac{L_b}{16\pi^2}\, ,\\ \nonumber
\Lambda_6(\mu)&=&\lambda_6T
-T\left[12\lambda_1\lambda_6
+3\lambda_3\lambda_6
+3\lambda_3\lambda_7
+4\lambda_4\lambda_6
+2\lambda_4\lambda_7
+10\lambda_5\lambda_6
+2\lambda_5\lambda_7\right.\\&&
\left.\null -\frac{9}{2}\lambda_6g^2\right]
\frac{L_b}{16\pi^2}\, ,\\ \nonumber
\Lambda_7(\mu)&=&\lambda_7T
-T\left[12\lambda_2\lambda_7
+3\lambda_3\lambda_6
+3\lambda_3\lambda_7
+2\lambda_4\lambda_6
+4\lambda_4\lambda_7 
+2\lambda_5\lambda_6
+10\lambda_5\lambda_7
\right.
\\&&
\left.\null -\frac{9}{2}\lambda_7g^2\right]
\frac{L_b}{16\pi^2}\, .
\eqa
The coupling constants $g^2_E(\mu)$, $h^2_E(\mu)$ and $\Lambda^4_E(\mu)$
have been computed in e.g. Refs.~\cite{laine,marta}:
\bqa
g_E^2(\mu)&=&g^2T\left[1+\frac{g^2}{16\pi^2}\left(7
L_b+\frac{2}{3}\right)\right]\, ,\\
h_E^2(\mu)&=&{1\over4}g^2T\left[1+\frac{g^2}{16\pi^2}
\left(7
L_b
+\frac{49}{6}\right)+\frac{3\lambda_1}{4\pi^2}
+\frac{\lambda_3}{4\pi^2}+\frac{\lambda_4}{8\pi^2}\right]\, ,\\
\Lambda_E(\mu)&=&\frac{3g^4T}{8\pi^2}\, .
\eqa
The coupling constants are all renormalization group invariant to this
order, which can be verified by using the renormaliztion group
equations. This property holds to all orders in perturbation theory
if the effective Lagrangian is restricted to  
superrenormalizable terms.

The scalar mass parameters have been computed in the one-loop
approximation by Losada in Ref.~\cite{marta}.
Here, we present results for the mass parameters to two-loop order:
\bqa\nonumber
\label{massen}
M_1^2(\mu)&=&m_1^2-\left[6m_1^2\lambda_1+2m_2^2\lambda_3 +m_2^2\lambda_4
+6m^2_3\lambda_6-\frac{9}{4}m_1^2g^2\right]\frac{L_b}{16\pi^2}\\ \nonumber
&&
\null +[6\Lambda_1+2\Lambda_3+\Lambda_4+\frac{9}{4}g^2_E]\frac{T^2}{12}
+\frac{T^2}{16\pi^2}\left[\frac{3}{4}\lambda_1g^2+\frac{1}{4}\lambda_3g^2
+\frac{1}{8}\lambda_4g^2+\frac{45}{32}g^4\right]
\\ \nonumber
&&\null-{T^2\over16\pi^2}\left[12\lambda_1^2+2\lambda_3^2+2\lambda_3\lambda_4+
2\lambda^2_4+12\lambda_5^2
+9\lambda_6^2+3\lambda_7^2-9\lambda_1g^2-3\lambda_3g^2
\right.\\
&&
\left.\null-\frac{3}{2}\lambda_4g^2
-\frac{75}{16}g^4\right]\left[\ln\frac{3T}{\mu}+c\right]\, , \\  \nonumber
M_2^2(\mu)&=&m_2^2-\left[6m_2^2\lambda_2+2m_1^2\lambda_3 +m_1^2\lambda_4
+6m^2_3\lambda_7-\frac{9}{4}g^2m_2^2\right]\frac{L_b}{16\pi^2} \\ \nonumber
&&
\null +(6\Lambda_2+2\Lambda_3+\Lambda_4+\frac{9}{4}g^2_E)\frac{T^2}{12}
+\frac{T^2}{16\pi^2}\left[\frac{3}{4}\lambda_2g^2+\frac{1}{4}\lambda_3g^2
+\frac{1}{8}\lambda_4g^2+\frac{45}{32}g^4\right]
\\ \nonumber
&&\null -{T^2\over16\pi^2}\left[12\lambda_2^2+2\lambda_3^2+2\lambda_3\lambda_4+2\lambda^2_4
+12\lambda_5^2
+3\lambda_6^2+9\lambda_7^2-9\lambda_2g^2-3\lambda_3g^2
\right.\\
&&
\left.\null-\frac{3}{2}\lambda_4g^2
-\frac{75}{16}g^4\right]\left[\ln\frac{3T}{\mu}+c\right]\, , \\ \nonumber
M_3^2(\mu)
&=&m_3^2-\left[m_3^2\lambda_3+2m_3^2\lambda_4+6m_3^2\lambda_5+3m_1^2\lambda_6
+3m_2^2\lambda_7-\frac{9}{4}g^2m_3^2\right]\frac{L_b}{16\pi^2}\\ \nonumber
&&\null +(\Lambda_6+\Lambda_7)\frac{T^2}{4}
+\frac{T^2}{16\pi^2}\left[\frac{3}{8}\lambda_6g^2
+\frac{3}{8}\lambda_7g^2\right]
\\ \nonumber
&&\null -\frac{T^2}{16\pi^2}\left[6\lambda_1\lambda_6+6\lambda_2\lambda_7
+3\lambda_3\lambda_6+3\lambda_3\lambda_7
+3\lambda_4\lambda_6+3\lambda_4\lambda_7
+6\lambda_5\lambda_6
+6\lambda_5\lambda_7\right.\\
&&
\left.\null -{9\over2}\lambda_6g^2
-{9\over2}\lambda_7g^2
\right]\left[\ln\frac{3T}{\mu}+c\right]\, .
\eqa
Here, $c$ is the
constant~\cite{rummu}
\bqa
c=\frac{1}{2}\left[\ln\frac{8\pi}{9}
+\frac{\zeta^{\prime}(2)}{\zeta (2)}-2\gamma_E\right]
\approx-0.348725\, .
\eqa
Note that we have written our mass parameters in terms of the renormalization
group invariant couplings of the $3d$ theory. The remaining dependence
on $\mu$ reveals that scalar mass parameters depend explicitly on the
scale $\mu$. This dependence on the scale $\mu$ is canceled by the
scale dependence arising from calculations in the effective theory.

The Debye mass is normally needed in the one-loop approximation~\cite{laine}
\bqa
m_E^2(\mu)&=&g^2T^2\,.
\eqa
There is no dependence on $\mu$ at leading order in $g^2$.
\section{Integrating out $A_0^{a \prime}$}
The next step is to integrate out the adjoint scalar triplet $A_0^{a \prime}$.
This is carried out by calculating correlators in the two theories
at distances $R\gg 1/gT$ and require that they be the same~[27-28].
The parameters in the effective theory 
encode the physics on the scales $T$ and $gT$ and
are called middle-distance coefficients.
Before doing this, however, 
we must determine the masses of the scalar doublets
near the phase transition. This is done by constructing the scalar mass
matrix and finding the temperatures at which it has zero eigenvalues.
The higher of these temperatures is close to $T_c$, where the phase transition
takes place and the corresponding eigenvector (Higgs doublet)
has a mass of order $g^2T$. 
The mass of the second scalar multiplet
(after diagonalization) is determined near $T_c$
and it is found that it is generally of order $gT$ and it
should be integrated out together with $A_0^{a \prime}$~\cite{laine}. 
Only with fine-tuning
of the parameters in the 2HDM, is it possible to obtain a mass of order 
$g^2T$~\cite{laine}. 
In this case it must be kept in the second effective Lagrangian.
Both cases are considered below. The 
diagonalization modifies the parameters of~(\ref{effprim}) 
and the relations between the old and new parameters
can be found in~\cite{mikko,marta}. 
In the following it is the rotated parameters
of~(\ref{effprim}) that appear in the formulas.
\subsection{One Heavy Higgs and one Light Higgs}
In this subsection we consider the case where one of the Higgs fields 
(denoted by $\Phi_2^{\prime}$) is heavy and has a mass of order $gT$. 
Hence, we integrate out this field together with the adjoint Higgs field
$A_0^{a}$.
The second effective field theory is then $SU(2)$+one Higgs doublet
with higher order operators satisfying the symmetries.
The Lagrangian reads
\bqa
\tilde{{\cal L}}_{\mbox{\scriptsize eff}}&=&\frac{1}{4}\tilde{G_{ij}}
\tilde{G_{ij}}
+\tilde{M}_1^2(\mu)\tilde{\Phi}_1^{\dagger}\tilde{\Phi}_1
+(D_i\tilde{\Phi}_1)^{\dagger}(D_i\tilde{\Phi}_1)+
\tilde{\Lambda}_1(\mu)(\tilde{\Phi}_1^{\dagger}\tilde{\Phi}_1)^2
+\delta\tilde{{\cal L}}_{\mbox{\scriptsize eff}}\, .
\eqa 
The gauge coupling is denoted by $g_M^2(\mu)$.\\ \\
The field renormalization constant for the scalar fields 
vanish in the one-loop approximation, since there are no
momentum dependent one-loop diagrams with internal $A_0^{a \prime}$'s. Thus
\beq
\label{scalar}
\tilde{\Phi}_i(\mu)\approx\Phi_i^{\prime}(\mu)\, .
\eeq
This is in contrast with the gauge fields, since there is a momentum
dependent one-loop diagram with $A_0^{a \prime}$ on the internal lines
and $A_i^{a \prime}$ on the external lines. 
The result is~\cite{mikko}
\beq
\tilde{A}_i^{a}\approx A_i^{a \prime}\left[1+\frac{g^2_E}{24\pi m_E}
+\frac{g^2_E}{48\pi M_2}\right]^{1/2}\, .
\eeq
The results for the coupling constants can be found in~\cite{mikko,marta}:
\bqa
\tilde{\Lambda}_1(\mu)&=&\Lambda_1(\mu)-\frac{1}{16\pi M_2}
[2\Lambda_3^2+2\Lambda_3\Lambda_4
+\Lambda^2_4+4\Lambda_5^2+24\Lambda^2_6-24\Lambda_6\Lambda_7]
-\frac{3h_E^2}{8\pi m_E}\, ,\\
g_M^2(\mu)&=&g^2_E(\mu)\left[1-\frac{g^2_E}{24\pi m_E}
-\frac{g^2_E}{48\pi M_2}\right]\, .
\eqa
The mass parameter has previously been computed by Losada~\cite{marta}
at one-loop.
The result in the two-loop approximation is:
\bqa\nonumber
\tilde{M}^2_1(\mu)&=&M^2_1(\mu)
-{1\over4\pi}\left[\left(2\Lambda_3+\Lambda_4\right)M_2+3h_E^2m_E\right]
+\frac{3}{16\pi^2}\left[
2\Lambda_2\Lambda_3+\Lambda_2\Lambda_4
+6\Lambda_6\Lambda_7\right]\\ \nonumber
&&
\null -{9\over16\pi^2}\Lambda_6^2\left[\ln\frac{\mu}{M_2}+{1\over2}\right]
-\frac{3}{16\pi^2}\Lambda^2_7\left[\ln\frac{\mu}{3M_2}+\frac{1}{2}\right]
\\ \nonumber&&
\null-\frac{1}{16\pi^2}\left[\left(2\Lambda_3^2+2\Lambda_3\Lambda_4+2\Lambda^2_4
+12\Lambda_5^2-3\Lambda_3g_E^2-{3\over2}\Lambda_4g_E^2\right)
\ln\frac{\mu}{2M_2}\right. \\ \nonumber
&&
\left.\null+\Lambda_3^2+\Lambda_3\Lambda_4+\Lambda_2^4+6\Lambda_5^2
-{3\over4}\Lambda_3g_E^2-{3\over8}\Lambda_4g_E^2\right] \\ \nonumber
&&
\null -\frac{1}{16\pi^2}\left[\left(6h_E^4-12h_E^2g_E^2+{3\over4}g_E^4\right)
\ln\frac{\mu}{2m_E}+3h_E^4-3h_E^2g_E^2\right]
\\ &&
\null +\frac{1}{16\pi^2}\left[6h_E^4\frac{M_2}{m_E}+
3\Lambda_3h_E^2\frac{m_E}{M_2}+{3\over2}\Lambda_4h_E^2\frac{m_E}{M_2}\right]\,.
\eqa
\subsection{Two Light Higgs Doublets}
In this subsection we consider the other scenario when both 
Higgs fields have masses of order $g^2T$.
The effective Lagrangian is now a three-dimensional 2HDM
with additional higher order operators which satisfy the symmetries at
high temperature:
\bqa\nonumber
\tilde{{\cal L}}_{\mbox{\scriptsize eff}}&=&\frac{1}{4}\tilde{G_{ij}}
\tilde{
G_{ij}}
+(D_i\tilde{\Phi}_1)^{\dagger}(D_i\tilde{\Phi}_1)
+(D_i\tilde{\Phi}_2)^{\dagger}(D_i\tilde{\Phi}_2)
+\tilde{M}_1^2(\mu)\tilde{\Phi}_1^{\dagger}\tilde{\Phi}_1
+\tilde{M}_2^2(\mu)\tilde{\Phi}_2^{\dagger}\tilde{\Phi}_2\\
&&
\null+\tilde{M}_3^2(\mu)(\tilde{\Phi}^{\dagger}_1\tilde{\Phi}_2+
\tilde{\Phi}_2^{\dagger}\tilde{\Phi}_1)
+V(\tilde{\Phi}_1,\tilde{\Phi}_2)
+\delta\tilde{{\cal L}}_{\mbox{\scriptsize eff}}\, .
\eqa
The scalar couplings are denoted by $\tilde{\Lambda}_i$ and the 
gauge coupling by $g_M^2(\mu)$.\\ \\
Again the scalar fields are not renormalized by 
integrating out 
the $A_0^{a \prime}$ fields and so (\ref{scalar}) also holds in the present
case.
This is in contrast with the gauge fields, since there is a trilinear
coupling between $A_0^{a \prime}$ and $A_i^{a \prime}$:
\beq
\tilde{A}_i^{a}\approx A_i^{a \prime}\left[1+\frac{g^2_E}{24\pi m_E}\right]^{1/2}\, .
\eeq
The scalar couplings $\tilde{\Lambda}_1(\mu)$$-$$\tilde{\Lambda}_3(\mu)$ get modified
by the integrating out $A_0^a$:
\bqa
\tilde{\Lambda}_1(\mu)=\Lambda_1(\mu)-\frac{3h^4_E}{8\pi m_E}\, ,\hspace{0.8cm}
\tilde{\Lambda}_2(\mu)=\Lambda_2(\mu)-\frac{3h^4_E}{8\pi m_E}\, ,\hspace{0.8cm}
\tilde{\Lambda}_3(\mu)=\Lambda_3(\mu)-\frac{3h^4_E}{4\pi m_E}\, .
\eqa
The other coupling constants, $\tilde{\Lambda}_4(\mu)-\tilde{\Lambda}_7(\mu)$,
are not modified in this step.
The gauge coupling reads
\beq
g_M^2(\mu)=g^2_E(\mu)\left[1-\frac{g^2_E}{24\pi m_E}\right]\, .
\eeq
The expression for the mass parameters at the two-loop level are
\bqa
\tilde{M}_1^2(\mu)&=&M_1^2(\mu)-\frac{3h_E^2m_E}{4\pi}
-\frac{1}{16\pi^2}\left[\left(6h_E^4-12h_E^2g_E^2+{3\over4}g_E^4\right)
\ln\frac{\mu}{2m_E}+
3h_E^4-3h_E^2g_E^2\right]\, , \\
\tilde{M}_2^{2}(\mu)&=&M_2^2(\mu)
-\frac{3h_E^2m_E}{4\pi}
-\frac{1}{16\pi^2}\left[\left(6h_E^4-12h_E^2g_E^2+{3\over4}g_E^4\right)
\ln\frac{\mu}{2m_E}
+3h_E^4-3h_E^2g_E^2\right] \,, \\
\tilde{M}_3(\mu)&=&M_3(\mu)\,.
\eqa
\section{Summary}
In the present paper I have applied the effective field theory methods
developed in Refs.~[27-28] to the 2HDM.
I have exploited the fact that there are three well separated momentum scales
and constructed a sequence of two effective three-dimensional
field theories.
The parameters in the final effective Lagrangian have previosuly been
calculated in the one-loop approximation~\cite{marta}. The two-loop results
presented here are new.

The resulting field theory can be used for investigating several
aspects of the phase transition
in the Two Higgs Doublet Model. This includes in particular the 
strength of the phase transition, and also the sphaleron rate
immediately after the completion of the phase transition.  
\appendix\bigskip\renewcommand{\theequation}{\thesection.\arabic{equation}}
\setcounter{equation}{0}\section{Notation and Conventions}
Throughout the work we use the imaginary 
time formalism, where the four-momentum is $P=(p_{0},{\bf p})$
with $P^{2}=p_{0}^{2}+{\bf p}^{2}$. 
The Euclidean energy takes on discrete values, $p_{0}=2n\pi T$
for bosons. 
Dimensional regularization is used to
regularize both infrared and ultraviolet divergences by working
in $d=4-2\epsilon$ dimensions, 
and we apply the $\overline{MS}$ 
renormalization scheme. We shall use the following notations for the 
sum-integrals that appear
\bqa
\label{ana}
\hbox{$\sum$}\!\!\!\!\!\!\int_Pf(P)&\equiv &\left( \frac{e^{\gamma_{\tiny E}}\mu^{2}}
{4\pi}\right)^{\epsilon}\,\,\,\,
T\!\!\!\!\!
\sum_{p_{0}=2\pi nT}\int\frac{d^{3-2\epsilon}p}
{(2\pi)^{3-2\epsilon}}f(P)\, .
\eqa

The one-loop sum-integrals needed in this work have been calculated in 
e.g. Refs.~[8,28,31]. 
We list them here for the convenience of the reader:
\bqa
\hbox{$\sum$}\!\!\!\!\!\!\int_P\frac{1}{P^{2}}&=&\frac{T^{2}}{12}
\left[1+\epsilon l_{\epsilon}
\right]\, ,\\
\hbox{$\sum$}\!\!\!\!\!\!\int_P\frac{1}{(P^{2})^{2}}&=&\frac{1}{16\pi^{2}}
\left[\frac{1}{\epsilon}+L_b+{\cal O}(\epsilon)\right]\, ,\\
\hbox{$\sum$}\!\!\!\!\!\!\int_P\frac{P_{0}^{2}}{(P^{2})^{2}}&=&-\frac{T^{2}}{24}
\left[1+\epsilon (l_{\epsilon}-2)
\right]\,,\\
\hbox{$\sum$}\!\!\!\!\!\!\int_P\frac{P_{0}^{2}}{(P^{2})^{3}}&=&
\frac{1}{64\pi^{2}}
\left[\frac{1}{\epsilon}+L_b+2+{\cal O}(\epsilon)\right]\,,\\
\label{show}
\hbox{$\sum$}\!\!\!\!\!\!\int_P\frac{P_{0}^{4}}{(P^{2})^{4}}&=&
\frac{1}{128\pi^{2}}
\left[\frac{1}{\epsilon}+L_b+\frac{8}{3}+{\cal O}(\epsilon)\right].
\eqa
Here 
\bqa
L_b=2\ln\frac{\mu}{4\pi T}+2\gamma_E\, ,\hspace{1cm}
l_{\epsilon}=2\ln\frac{\mu}{T}+2\gamma_E-2\ln2
-2\frac{\zeta^{\prime}(2)}{\zeta (2)}\, .
\eqa
Moreover, $\gamma_E$ is the Euler-Mascharoni constant and $\zeta (x)$ is the
Riemann Zeta function.

In the calculation of the mass parameters, we also need the fact that 
two-loop setting sun diagram is zero~\cite{braaten}:
\beq
\hbox{$\sum$}\!\!\!\!\!\!\int_{PQ}\frac{1}{P^{2}Q^2(P+Q)^2}=0\,.
\eeq
In the effective three-dimensional theory we also 
use dimensional regularization
in $3-2\epsilon$ dimensions to regularize infrared and ultraviolet 
divergences.
In analogy with~(\ref{ana}), we define
\beq
\int_{p}f(p)\equiv\left( \frac{e^{\gamma_{\tiny E}}\mu^{2}}
{4\pi}\right)^{\epsilon}\int\frac{d^{3-2\epsilon}p}
{(2\pi)^{3-2\epsilon}}f(p)\,.
\eeq
Again $\mu$ coincides with the renormalization scale in the 
modified minimal subtraction renormalization
scheme.

The one-loop and two-loop integrals needed are
\bqa
\label{effloop}
\int_{p} \frac{1}{p^{2}+m^{2}}&=&-\frac{m}{4\pi}
\left[1+{\cal O}(\epsilon)\right]\, ,\\
\label{elgen}
\int_{p} \frac{1}{(p^{2}+m^{2}_1)(p^2+m_2^2)}&=&\frac{1}{4\pi (m_1+m_2)}\left[1+{\cal O}(\epsilon)\right]\, ,\\ \nonumber
\int_{pq} \frac{1}{(p^{2}+m^{2}_1)(q^{2}+m^{2}_2)[({\bf p}-{\bf q})^{2}+m_3^2]}
&=&\frac{1}{16\pi^2}
\left[\frac{1}{4\epsilon}
+\frac{1}{2}\right.\\
&&\left.+\ln\frac{\mu}{m_1+m_2+m_3}+{\cal O}(\epsilon)\right]\, ,\\
\int_{pq} \frac{1}{(p^{2}+m^{2})(q^2+m^2)^2({\bf p}-{\bf q})^{2}}
&=&\frac{1}{16\pi^2m^2}\left[{1\over4}+{\cal O}(\epsilon)\right]\, ,  \\
\int_{pq} \frac{1}{(p^{2}+m^{2})(q^2+m^2)({\bf p}-{\bf q})^{4}}
&=&\frac{1}{16\pi^2m^2}\left[-{1\over8}+{\cal O}(\epsilon)\right]\, .
\eqa
These integrals have been computed by several authors, e.g. in 
Refs.~\cite{arnold12,braaten,laine}.
\setcounter{equation}{0}
\section{Matching Example}
In this appendix we explicitly show how the matching procedure is carried
out by determining the mass parameter $M_1^2(\Lambda)$ to two-loop order.

We denote the static two-point
function of the Higgs field in the full theory by 
$\Gamma_{\phi_1,\phi_1}^{(2)}({\bf k})$, 
and the static two-point function in the effective
theory by 
$\Gamma_{\phi_1^{\prime},\phi_1^{\prime}}^{(2)}({\bf k})$. 
The corresponding self-energies are denoted by $\Sigma ({\bf k})$ and
$\tilde{\Sigma}({\bf k})$. Finally, the n'th order contribution to the
self-energies in the loop
expansion are denoted by $\Sigma^{(n)}({\bf k})$ and 
$\tilde{\Sigma}^{(n)}({\bf k})$.

The self-energies can be expanded in powers of the external momentum 
${\bf k}$ 
and so we can write the two-point functions as
\bqa
\Gamma_{\phi_1,\phi_1}^{(2)}({\bf k})&=&k^2+m_1^2+\Sigma^{(1)}(0)
+k^2\Sigma^{(1)\prime}(0)+\Sigma^{(2)}(0)\,,\\
\Gamma_{\phi_1^{\prime},\phi_1^{\prime}}^{(2)}({\bf k})&=&k^2+M_1^2(\mu)
+\tilde{\Sigma}^{(1)}(0)
+k^2\tilde{\Sigma}^{(1)\prime}(0)+\tilde{\Sigma}^{(2)}(0)+\delta M^2_1\, .
\eqa
Here, we have added a mass counterterm, which is associated with mass 
renormalization. 
The mass parameter is then
determined by matching these two-point functions, and by taking the  
field normalization constant into account, we can write the
matching equation as
\beq
\label{match}
\Gamma_{\phi_1,\phi_1}^{(2)}({\bf k})=\left[1+\Sigma^{(1)\prime}(0)\right]
\Gamma_{\phi_1^{\prime},\phi_1^{\prime}}^{(2)}({\bf k})\, .
\eeq
Since the external momentum ${\bf k}$
provides the only mass scale in the loop integrals contributing
to the self-energy of the effective theory, they all vanish
in dimensional regularization. The matching equation~(\ref{match})
can then be rewritten as 
\beq
M_1^2(\mu)=m_1^2\left[1-\Sigma^{(1)\prime}(0)\right]
+\Sigma^{(1)}(0)\left[1-\Sigma^{(1)\prime}(0)\right]
+\Sigma^{(2)}(0)-\delta M_1^2\,.
\eeq
The self-energy at one-loop order in the
full theory reads:
\bqa\nonumber
\Sigma^{(1)} ({\bf k})&=&
-\left[6m_1^2\lambda_1+2m_2^2\lambda_3+m_2^2\lambda_4+6m_3^2\lambda_6\right]
\hbox{$\sum$}\!\!\!\!\!\!\int_P\frac{1}{P^{4}}\\ \nonumber
&&\null+
\left[6\lambda_1+2\lambda_3+\lambda_4+\frac{3}{4}(d-1)g^2\right]
\hbox{$\sum$}\!\!\!\!\!\!\int_P\frac{1}{P^{2}}\\
&&
\null-3g^2\hbox{$\sum$}\!\!\!\!\!\!\int_P
\frac{k^2}{P^2(P+K)^2}
+3g^2\hbox{$\sum$}\!\!\!\!\!\!\int_P\frac{({\bf p}{\bf k})^2}{P^4(P+K)^2}\,.
\eqa
Expanding in powers of the external momentum ${\bf k}$ gives
\bqa\nonumber
\label{avles}
\Sigma^{(1)} ({\bf k})&=&
-\left[6m_1^2\lambda_1+2m_2^2\lambda_3+m_2^2\lambda_4+6m_3^2\lambda_6\right]
\hbox{$\sum$}\!\!\!\!\!\!\int_P\frac{1}{P^{4}}\\ 
&&
\null+\left[6\lambda_1+2\lambda_3+\lambda_4+\frac{3}{4}(d-1)g^2\right]
\hbox{$\sum$}\!\!\!\!\!\!\int_P\frac{1}{P^{2}}
-\frac{9}{4}g^2k^2\hbox{$\sum$}\!\!\!\!\!\!\int_P\frac{1}{P^4}       
+{\cal O}(k^4/T^2)\,.
\eqa
This implies
\bqa\nonumber
\label{sjoel1}
\Sigma^{(1)} (0)&=&
-\left[6m_1^2\lambda_1+2m_2^2\lambda_3+m_2^2\lambda_4+6m_3^2\lambda_6\right]
\hbox{$\sum$}\!\!\!\!\!\!\int_P\frac{1}{P^{4}}\\ &&
\null+\left[6\lambda_1+2\lambda_3+\lambda_4+\frac{3}{4}(d-1)g^2\right]
\hbox{$\sum$}\!\!\!\!\!\!\int_P\frac{1}{P^{2}}\,,\\ 
\label{sjoeld}
\Sigma^{(1)\prime} (0)&=&
-\frac{9}{4}g^2\hbox{$\sum$}\!\!\!\!\!\!\int_P\frac{1}{P^4}\,.                   
\eqa
The two-loop contribution to the self-energy at zero external momentum is
\bqa\nonumber
\Sigma^{(2)}(0)&=&-
\left[36\lambda_1^2
+12\lambda_1\lambda_3+6\lambda_1\lambda_4
+12\lambda_2\lambda_3+6\lambda_2\lambda_4+4\lambda^2_3
+4\lambda_3\lambda_4
+\lambda_4^2
\right.\\ \nonumber
&&\null +18\lambda_6^2+18\lambda_6\lambda_7
+\frac{9}{2}(d-1)\lambda_1g^2+\frac{3}{2}(d-1)\lambda_3g^2
+\frac{3}{4}(d-1)\lambda_4g^2 \\
&&\left.\null+\frac{1}{2}(3d^2-9d+6)g^4\right]
\hbox{$\sum$}\!\!\!\!\!\!\int_{PQ}\frac{1}{P^{2}Q^{4}}\, .
\eqa
After renormalization of the mass parameter $m_1^2$
as well as the coupling constants,
we are left with a pole in $\epsilon$.
This pole is canceled by the mass renormalization counterterm, which
is 
\bqa\nonumber
\delta M^2_1&=&\left[12\lambda_1^2+2\lambda_3^2+2\lambda_3\lambda_4+
2\lambda^2_4+12\lambda_5^2
+9\lambda_6^2+3\lambda_7^2-9\lambda_1g^2-3\lambda_3g^2\right.
\\ &&
\left.\null-\frac{3}{2}\lambda_4g^2
-\frac{75}{16}g^4\right]\frac{1}{64\pi^2\epsilon}\, .
\eqa
This is the result for the mass counterterm of the three-dimensional
2HDM at next-to-leading order in the coupling constants.
The mass parameter $M_1^2(\mu)$ is then given by (\ref{massen}).

This work was supported in part by a Faculty Development Grant from
the Physics Department of the Ohio State
University and by the Norwegian 
Research Council (project 124282/410).

\end{document}